\documentstyle[epsf]{article}%
\begin{document}
\title{A protein structural alphabet and its substitution matrix CLESUM}
\author{Wei-Mou Zheng${^{1,3}}$, Xin Liu${^2}$\\
\small ${^1}${\it Institute of Theoretical Physics, Academia
Sinica, Beijing 100080, China}\\
\small ${^2}${\it The Interdisciplinary Center of Theoretical
Studies, Academia Sinica, Beijing 100080, China}\\
\small ${^3}$ To whom correspondence should be addressed. }
\date{}
\maketitle

\begin{abstract}
By using a mixture model for the density distribution of the three pseudobond
angles formed by $C_\alpha$ atoms of four consecutive residues, the local
structural states are discretized as 17 conformational letters of a protein
structural alphabet. This coarse-graining procedure converts a 3D structure to a 1D
code sequence. A substitution matrix between these letters is constructed
based on the structural alignments of the FSSP database.
\end{abstract}

\noindent Key words: structure alignment; structural codes; structural
substitution matrix.
%\leftline{PACS number(s): 87.10.+e,02.50.-r}
\bigskip

\section{Introduction}
Drastic approximations are unavoidable in prediction of protein structure %tertiary
from the amino acid sequence. Most local structure prediction
methods use three secondary structure states: helix, strand and
loop. However, segments of a single secondary structure may vary
significantly in their 3D structures. A refined objective
classification of segments may enhance our ability in the prediction
of structures, and deepen our understanding of the modular
architecture of proteins.

The usual approaches simplify protein structure by modelling
proteins as chains of one or two interacting centers representing
individual amino acids, and adopt only a small number of discrete
conformational states. Many studies to investigate the
classification of protein fragments use the backbone $(\phi , \psi
)$ dihedral angles, or angles of $C_\alpha$ psuedobonds or distances derived
from the positions of $C_\alpha$ atoms. Due to the anticorrelation
between $\phi$ and $\psi$ (McCammon et al., 1977; Flocco and
Mowbray 1995), there may be instances where a big change in both
$\phi$ and $\psi$ does not represent an obvious change in the
$C_\alpha$ pseudobond angles, but a reorientation of the peptide
in question. Furthermore, the relation between $C_\alpha$
coordinates and pseudobond angles is rather straightforward, and
pseudobond angles have a more direct geometric meaning than
distances. We shall use only pseudobond angles in this paper.

By restricting the local conformations of individual residues to a
handful of states, one can discretize protein conformation to
convert the 3D structure of a backbone to a 1D sequence of these
discrete states akin to the amino acid sequence. Prediction of
protein structure depends on the accuracy and complexity of the
models used. A model must be as simple as possible to reduce the
conformational space to be searched for a ¡®¡®correct¡¯¡¯
conformation, while a model of low complexity tends to have a lower
accuracy. A model must represent the actual geometry of protein
conformations accurately enough, but a complex model is prone to
over-fitting the observed data.

Generally, the procedure to deduce finite discrete conformational
states from a continuous conformational phase space is a
clustering analysis. There have been a variety of different ways
of clustering. For example, Park and Levitt (1995) represent the
polypeptide chain by a sequence of rigid fragments that are chosen
from a library of representative fragments, and concatenated
without any degrees of freedom. The average deviation of the
global-fit approximations over the training set is taken as the
objective function for optimizing the finite representative
fragments. The state clusters there are representative points of the phase
space. Rooman, Kocher and Wodak (1991) intuitively divide the $\phi$-$\psi$
space into 6 regions, which corresponds to a partitioning based on the
Ramachandran plot. Standard methods for clustering analysis have
been also used to generate discrete structure states (Bystroff and Baker, 1998).

Hidden Markov models (HMMs; Rabiner, 1989), possessing a rigorous
but flexible mathematical structure, have been used in a variety
of computational biology problems such as sequence motif
recognition (Fujiwara et al., 1994), gene finding (Burge and Karlin,
1997), protein secondary structure prediction (Asai, Hazamizu and Handa,
1993; Zheng, 2004), and multiple sequence alignments (Krogh {\it et al.},
1994). The HMMs have been also used for
identifying the modular framwork for the protein backbone
(Edgoose, Allison and Dowe, 1998; Camproux {\it et al.}, 1999). In
these HMMs conformation states are represented by probability
distributions, which is much finer than a simple partition of the
phase space. HMMs also take into account the sequential
connections between conformational states, hence involve in a
large number of parameters, which make the model training a tough
task. Furthermore, it is not so convenient to assign structure
codes to a short segment with HMMs.

Here we develop a description of protein backbone tertiary
structure using psuedobond angles of successive $C_\alpha$ atoms.
Finite conformational states as structural alphabet are selected
according to the density peaks of probability distribution in the
phase space spanned by pseudobond angles, and their feasibility
of characterizing short segment polypeptide backbone conformation
is examined. In order to use the structural codes in the structural
comparison, we derive a substitution matrix of these conformational
states from a representative pairwise aligned structure set of the
FSSP (families of structurally similar proteins) database of Holm
and Sander (1994).

\section{Methods}
 Among a variety of abstract representing forms for protein 3D structure,
a frequently encountered one is the protein virtual backbone. The
$C_\alpha$ atom of the residue is chosen as the representative
point. In this representation, two adjacent residues in a protein
sequence are virtually bonded, forming a pseudobond.

\subsection{Pseudo-bond angles}
The virtual bond bending angle $\theta$ defined for three
contiguous points $(a, b, c)$ is the angle between the vectors
${\bf r}_{ab}={\bf r}_b-{\bf r}_a$ and ${\bf r}_{bc}$, i.e.
$\theta ={\bf r}_{ab}\cdot
{\bf r}_{bc}/(|r_{ab}r_{bc}|)$. The range of $\theta$ is $[0, 2\pi
]$. The virtual bond torsion angle $\tau$ defined for four
contiguous points $(a, b, c, d)$ is the dihedral angle between the
planes $abc$ and $bcd$. The range of $\tau$ is $(-\pi ,\pi ]$, and
its sign is the same as $({\bf r}_{ab}\times {\bf r}_{bc})\cdot
{\bf r}_{cd}$. In fact, we may adopt a wider range of $\tau$ under
the equivalence relation that $\tau_1$ and $\tau_2$ are equivalent
if $\tau_1= \tau_2 \ ({\rm mod}\  2\pi )$. For the four-residue
segment $abcd$, by takeing $a$ as the origin, and $b$ on the
$x$-axis, and $c$ on the $xy$-plane, the number of independent
relative coordinates are 6. The assumption of the fixed pseudobond
length, which is 3.8 \AA\ for the dominating {\it trans} peptide,
further reduces the number of degrees of freedom to 3. These independent
coordinates correspond to the angles $(\theta_{abc}, \tau_{abcd},
\theta_{bcd})$. Elongating the segment by one residue $e$ will add
two more angles $\tau_{bcde}$ and $\theta_{cde}$. Generally, for a
sequence of $n$ residues, we have $n-2$ bending angles and $n-3$
torsion angles, $2n-5$ in total. We shall assign the angle pair
$(\tau_{abcd}, \theta_{bcd})\equiv (\tau_c, \theta_c)$ to residue
$c$, the third of the four-residue segment.

Bending and torsion angles of a chain correspond to curvature and
torsion of a curve. The relative coordinates of the chain $\{
{\bf r}_0, {\bf r}_1,\cdots {\bf r}_n\}$ can be recovered from
their $2n-5$ angles $\{\theta_1;\tau_2, \theta_2; \cdots
;\tau_{n-1}, \theta_{n-1}\}$. By convention, we set the origin at
${\bf r}_0$, put ${\bf r}_1$ along the $x$-axis, and add
$\tau_1=0$. Introducing the rotation matrices $R_\theta$ and
$R_\tau$ (with respect to the $z$- and $x$-axis, respectively)
\begin{equation}
R_\theta = \left( \begin{array}{ccc} \cos\theta & -\sin\theta &0\\
\sin\theta &\cos\theta & 0\\ 0& 0& 1\end{array}\right), \qquad
R_\tau = \left( \begin{array}{ccc} 1& 0& 0\\ 0& \cos\tau & -\sin\tau \\
0& \sin\tau &\cos\tau \end{array}\right), {\quad\rm and \quad}
{\bf d} ={\bf r}_1 =\left(\begin{array}{c} 1\\ 0\\
0\end{array}\right) ,
\end{equation}
position ${\bf r}_k$ is determined by
\begin{equation}
T_0=I,\quad {\bf r}_0=0\cdot {\bf d},\quad T_k=T_{k-1} R_{\tau_k}R_{\theta_k},
\quad {\bf d}_k=T_{k-1}\cdot {\bf d}, \quad {\bf r}_k = {\bf r}_{k-1}+{\bf
d}_k , \quad k\geq 1,
\end{equation}
where $I$ is the identity matrix.

Longer fragments will include more correlation than shorter fragments. However,
the complexity that can be explored with the longer fragment lengths
is limited severely by the relatively small number of known protein structures,
and a larger number of discrete states have to be determined for a longer segment.
The minimal unit where the relative coordinates fix the angles and vice versa is
four contiguous residue segment. We shall concentrate mainly on the structure
codes for the four residue unit.

\subsection{The mixture model for the angle probability distribution}
The three pseudobond angles $(\theta ,\tau ,\theta' )$ of the
four-residue unit span the three-dimensional phase space. Our
classifiers for conformational states are based on the following
mixture model $M$: The probability distribution of `points' ${\bf
x}\equiv (\theta ,\tau ,\theta' )$ is given by the mixture of
several normal distributions
\begin{equation}
P({\bf x}|M)= \sum_{i=1}^c \pi_i N({\bf \mu}_i,{\bf \Sigma}_i),
\end{equation}
where $c$ is the number of the normal distribution categories in the mixture,
$\pi_i$ the prior for category $i$, and $N({\bf \mu},{\bf \Sigma})$ the normal
distribution
\begin{equation}
N({\bf \mu},{\bf \Sigma}) = (2\pi)^{-3/2}|{\bf \Sigma}|^{-1/2}
\exp [ \hbox{$\frac 12$}({\bf x}-{\bf \mu})\cdot {\bf \Sigma}^{-1}\cdot ({\bf
x}-{\bf \mu})].
\end{equation}
Each normal distribution has 6 parameters for its symmetric covariance matrix
${\bf \Sigma}$ and three for its mean ${\bf \mu }$. Adding one more parameter of
the prior for each category, the mixture model has 10$c$ parameters for the total
$c$ categories. (The normalization $\sum_i \pi_i =1$ reduces the number to
$10c -1$.) These categories will be translated as the structure codes.

To objectively determine the number $c$ of categories, we
investigate density peaks in the phase space with the downhill
simplex method of Nelder and Mead (1965). The method requires only
function evaluations, not derivatives. It is not very efficient in
terms of the number of function evaluations that it requires, but
still works well for our problem here. We use counts in a
rectangular box as the value of the function for optimization at
the center of the box. The box size corresponds to the Parzon window
width. A large box size has a low resolution, hence help us to
focus on main density peaks in the phase space, and to easily
locate them near their real location. Reducing the box size, we
can see more peaks which are less conspicuous and then unseen
under a larger box size. A too small box size, making local
fluctuations visible, is often misleading. Missing out any
important modes will affect the model training and the efficacy of
the structural codes generated. We first search for maximal points
of the one-dimensional marginal probability distributions of
$\theta$ and $\tau$, and then utilize them to generate a grid in
the $(\theta ,\tau , \theta ')$ space for searching for peaks in
the space.
%Nelder, J.A., and Mead, R. 1965, Computer Journal, vol. 7, pp. 308¨C313.

We examine also density peaks in the five-dimensional phase space spanned by
$(\theta_b,\tau_c,\theta_c, \tau_d,\theta_d)$ of the five-residue unit $abcde$
to investigate the effect of the angle correlation. A five-angle mode
$(\theta_b,\tau_c,\theta_c, \tau_d,\theta_d)$ contains two three-angle modes
$(\theta_b,\tau_c,\theta_c)$ and $(\theta_c, \tau_d,\theta_d)$. It is demanded
that all the important three-angle modes implied by the main density peaks in the
five-angle phase space must be included in the modes used for the construction
of the mixture model.

The main purpose of searching for density peaks is to estimate
the number $c$ of categories and $\{ {\bf
\mu}_i\}$ for each category. Once this has been done, we may start with some simple
$\{ \pi_i\}$ and $\{ {\bf \Sigma}_i\}$, say $\pi_i=1/c$ and certain diagonal
$\{ {\bf \Sigma}_i\}$, and then update the mixture model by the
Expectation-Maximization (EM) method as follows. For each point ${\bf x}_k =
(\theta_{k-1} ,\tau_k , \theta_k)$, we calculate the probability for the point to
belong to the $i$-th category $C_i$ according to the Bayes formula as
\begin{eqnarray}
P(C_i|{\bf x}_k) &\propto & \pi_i P({\bf x}_k|C_i) \nonumber \\
&\propto & \pi_i |{\bf \Sigma}_i|^{-1/2} \exp [\hbox{$\frac 12$} ({\bf x}_k-{\bf
\mu}_i)\cdot {\bf \Sigma}^{-1}_i\cdot ({\bf x}_k-{\bf \mu}_k)],
\end{eqnarray}
where we always shift $\tau_k$ to the interval $[\tau^{(i)}-\pi
,\tau^{(i)}+\pi )$ centered at $\tau^{(i)}$ of the $\tau$-component
of the mean ${\bf \mu}_i$. The probability $P(C_i|{\bf x}_k)$
satisfies the normalization condition $\sum_{i=1}^c P(C_i|{\bf x}_k)
=1$. The updated parameters for the mixture model are
estimated by the EM method as
\begin{eqnarray}
n_i &=& \sum_k P(C_i|{\bf x}_k), \qquad \pi_i=n_i/n, \quad n=\sum_i n_i, \\ %\nonumber \\
{\bf \mu}_i&=& \frac 1 n_i\sum_k P(C_i|{\bf x}_k) {\bf x}_k,  \\
{\bf \Sigma}_i &=& \sum_k P(C_i|{\bf x}_k) ({\bf x}_k -
{\bf\mu}_k) ({\bf x}_k - {\bf\mu}_k)^T. %\nonumber
\end{eqnarray}

Generally, the objective function for optimizing the mixture model is
\begin{equation}
{\rm Prob} (\{{\bf x}_k\}) =\prod_k\, \sum_i P({\bf x}_k,C_i)\propto
\prod_k\, \sum_i P(C_i|{\bf x}_k).
\end{equation}
However, when we convert point ${\bf x}_k$ to its structural code $i^*$, we use
\begin{equation}
i^* = {\rm arg}_i \max P(C_i|{\bf x}_k).
\label{i*}\end{equation}
An alternative objective function would be
\begin{equation}
Q(\{{\bf x}_k\}) =\prod_k\, \max_i P(C_i|{\bf x}_k).
\end{equation}
When starting with narrow distributions for ${\bf \Sigma}_i$, a very high value
of $Q$ could be seen at the first step. However, by just one step of the EM
iteration $Q$ will drop significantly, and then increases at later steps. While
${\rm Prob} (\{{\bf x}_k\})$ never decreases, $Q$ will decrease after reaching its
maximum. We may stop the model training before $Q$ decreases again. Thus, the
optimization here is a compromise between Prob$ (\{{\bf x}_k\})$ and
$Q(\{{\bf x}_k\})$.

Once we have the model, we may convert a structure to its conformational code sequence
according to (\ref{i*}). Although no effect from the connection of states is directly
considered, the model gains the advantage in being able to easily assign codes to
short fragments.

\begin{center}
\parbox{15cm}{Table 1. The 17 structural states from the mixture
model.}\end{center}%\smallskip

\begin{tabular}{c|r|r|rrr|rrrrrr}\hline
&\multicolumn{1}{c|}{$\pi$}&\multicolumn{1}{c|}{$|\Sigma |^{-1/2}$}
&\multicolumn{3}{c|}{ ${\bf \mu}$ } &\multicolumn{6}{c}{${\bf\Sigma}^{-1}$}\\
State&& &\multicolumn{1}{c}{$\theta$}& \multicolumn{1}{c}{$\tau$}&
\multicolumn{1}{c|} {$\theta '$}&
\multicolumn{1}{c}{$\theta\theta$}&
\multicolumn{1}{c}{$\tau\theta$} & \multicolumn{1}{c}{$\tau\tau$}&
\multicolumn{1}{c}{$\theta '\theta$}
& \multicolumn{1}{c}{$\theta '\tau$}& \multicolumn{1}{c}{$\theta '\theta '$}\\
\hline
I&  8.2& 1881& 1.52&  0.83&  1.52& 275.4& -28.3&  84.3& 106.9& -46.1& 214.4\\ %(a1
J&  7.3& 1797& 1.58&  1.05&  1.55& 314.3& -10.3&  46.0&  37.8& -70.0& 332.8\\ %a2
H& 16.2&10425&1.55&  0.88&  1.55& 706.6& -93.9& 245.5& 128.9& -171.8&786.1\\ %a1
K&  5.9&  254&  1.48&  0.70&  1.43&  73.8& -13.7&  21.5&  15.5& -25.3&  75.7\\ %a'
F&  4.9&  105&  1.09& -2.72&  0.91&  24.1&   1.9&  10.9& -11.2&  -8.8&  53.0\\ %(b1
E& 11.6&  109&  1.02& -2.98&  0.95&  34.3&   4.2&  15.2&  -9.3& -22.5&  56.8\\ %b1
C&  7.5&  100&  1.01& -1.88&  1.14&  28.0&   4.1&   6.2&   2.3&  -5.1&  69.4\\ %b2
D&  5.4&   78&  0.79& -2.30&  1.03&  56.2&   3.8&   4.2& -10.8&  -2.1&  30.1\\ %(b2
A&  4.3&  203&  1.02& -2.00&  1.55&  30.5&   9.1&   8.7&   6.0&   5.7& 228.6\\ %rb
B&  3.9&   66&  1.06& -2.94&  1.34&  26.9&   4.6&   4.9&   9.5&  -5.0&  54.3\\ %r2
G&  5.6&  133&  1.49&  2.09&  1.05& 163.9&   0.6&   3.8&   2.0&  -3.7&  32.3\\ %ra
L&  5.3&   40&  1.40&  0.75&  0.84&  43.7&   2.5&   1.4&  -7.0&  -2.9&  34.5\\ %(ra
M&  3.7&  144&  1.47&  1.64&  1.44&  72.9&   2.1&   4.8&   1.9&  -7.9&  72.9\\ %a+
N&  3.1&   74&  1.12&  0.14&  1.49&  25.3&   3.2&   3.1&   9.9&   0.9&  83.0\\ %r1
O&  2.1&  247&  1.54& -1.89&  1.48& 170.8&  -0.7&   3.7&  -4.1&   3.1&  98.7\\ %a-
P&  3.2&  206&  1.24& -2.98&  1.49&  48.0&   8.2&   7.3&  -4.9&  -6.6& 155.6\\ %(r2
Q&  1.7&   25&  0.86& -0.37&  1.01&  28.4&   1.5&   1.2&   3.4&   0.1&  19.5\\ %r'
\hline\end{tabular}%\smallskip

\section{Result}

For establishing the discrete structural states by training the
mixture model, we create a nonredundant set of 1544 non-membrane
proteins from PDB\_SELECT (Hobohm and Sander, 1994) with amino acid
identity less than 25\% issued on 25 September of 2001. The data of the
three-dimensional structures for these proteins are taken from
Protein Data Bank (PDB). The secondary structures for
these sequences are taken from the DSSP database (Kabsch and Sander, 1983). We
consider the reduced 3 secondary structure states $\{h, e, c\}$
generated from the 8 states of the DSSP by the coarse-graining
$H,G,I\to h$, $E\to e$ and $X,T,S,B\to c$. The total number of
contiguous fragments is 2248, which gives totally 264,232 points
in the three-angle phase space.

\begin{center}
\parbox{15cm}{Table 2. The percentages of each secondary structure in the structural states.\\
}\end{center}\vspace{-.5cm}{\small

\begin{tabular}{c rrrrrrrrrrrrrrrrrr}\hline
 &     $I$& $J$& $H$& $K$& $F$& $E$& $C$& $D$& $A$& $B$& $G$& $L$& $M$& $N$& $O$& $P$& $Q$&Counts\\
\hline
$cccc$&  3&   3&   1&   5&   3&   7&  14&   3&   7&   3&  10&   9&   5&   7&   4&   5&   3& 25090\\
$ccce$&  0&   1&   0&   3&   3&   6&  15&   2&   5&   4&  24&   9&   4&   4&   4&   4&   3&  3272\\
$ccch$&  2&   1&   1&   6&   5&   8&  22&   4&   3&   2&   9&  11&   7&   4&   1&   3&   2&  3028\\
$ccee$&  0&   0&   0&   0&   7&  20&  17&   7&   0&   1&  15&  24&   0&   0&   0&   0&   2&  4029\\
$cchh$&  1&   3&   0&   1&   0&   0&   1&   0&  46&   1&   0&   0&  17&   4&   0&  18&   1&  3664\\
$ceec$&  0&   0&   0&   0&   6&  36&  26&  14&   0&   8&   0&   2&   0&   0&   0&   0&   3&   620\\
$ceee$&  0&   0&   0&   0&   7&  43&  12&  22&   0&   4&   0&   2&   0&   0&   0&   2&   3&  3676\\
$ceeh$&  0&   0&   0&   0&   3&  11&  49&  19&   0&   3&   0&   0&   0&   0&   1&   1&   7&    51\\
$chhh$& 21&  38&  28&   4&   0&   0&   0&   0&   0&   0&   0&   0&   4&   2&   0&   0&   0&  4353\\
$eccc$&  3&   3&   1&   3&   2&   6&  15&   3&  14&   3&   5&  11&   4&   7&   2&   7&   4&  3007\\
$ecce$&  3&   1&   1&   6&   1&   4&   5&   1&   1&   1&   1&   4&   1&  26&  35&   0&   1&   492\\
$ecch$&  1&   1&   0&   5&   1&   6&  18&   5&   4&   1&   9&  24&   5&   6&   0&   3&   2&   258\\
$ecee$&  1&   0&   0&   0&   5&  18&  16&  10&   1&   2&   3&  33&   1&   0&   0&   2&   3&    80\\
$echh$&  0&   1&   0&   0&   0&   0&   0&   0&  52&   1&   1&   0&  10&   5&   0&  17&   5&   256\\
$eecc$&  0&   0&   0&   0&   6&  16&  19&   7&  12&   6&   0&   3&   0&   9&   0&  11&   4&  3807\\
$eece$&  0&   0&   0&   0&   6&  18&  21&  11&  11&   6&   0&   3&   1&  11&   0&   6&   2&    80\\
$eech$&  0&   0&   0&   0&   4&  15&  25&  16&   2&   5&   0&   3&   0&   6&   0&  10&  10&   256\\
$eeec$&  0&   0&   0&   0&   7&  36&  19&  14&   1&   9&   0&   6&   0&   0&   0&   1&   3&  3596\\
$eeee$&  0&   0&   0&   0&   5&  48&   8&  17&   0&   6&   0&   7&   0&   0&   0&   2&   1& 11418\\
$eeeh$&  0&   0&   0&   0&   4&  15&  41&  17&   2&  11&   0&   2&   0&   0&   0&   1&   4&   197\\
$eehh$&  0&   0&   0&   0&   0&   0&   3&   0&  57&   2&   0&   0&   0&   2&   1&  28&   4&   248\\
$ehhh$& 13&  43&  25&   3&   0&   0&   0&   0&   0&   0&   0&   0&   6&   5&   1&   0&   0&   248\\
$hccc$&  4&   5&   2&   4&   2&   4&   7&   1&   4&   2&  14&   6&   6&   5&  18&   6&   1&  3254\\
$hcce$&  1&   2&   0&   3&   5&   5&  11&   0&   7&   5&  22&   7&   4&   4&   6&   7&   3&   208\\
$hcch$&  3&   0&   1&   4&   3&   5&  21&   5&   4&   2&  14&   9&   7&   4&   4&   3&   1&   328\\
$hcee$&  0&   0&   0&   0&   6&  21&  18&   9&   0&   4&  17&  14&   1&   0&   0&   0&   2&   151\\
$hchh$&  1&   2&   0&   1&   0&   0&   1&   0&  12&   1&   0&   0&  27&  18&   0&  31&   0&   356\\
$heec$&  0&   0&   0&   0&   5&  50&  15&  10&   0&  10&   0&   5&   0&   0&   0&   5&   0&    20\\
$heee$&  0&   0&   0&   0&   4&  50&   4&  17&   1&   6&   0&   3&   0&   0&   0&   8&   1&   117\\
$hhcc$&  9&  15&  11&   9&   1&   0&   0&   0&   1&   0&  15&  10&   9&   1&  10&   2&   0&  3861\\
$hhce$&  1&   1&   0&   1&   2&   0&   2&   0&   1&   1&  43&  15&   6&   0&  14&   3&   0&   151\\
$hhch$&  8&   4&   3&  19&   3&   0&   1&   0&   0&   0&  15&  28&   7&   2&   3&   0&   0&   356\\
$hhee$&  0&   0&   0&   0&   0&   0&   0&   0&   0&   0&  52&  40&   3&   0&   0&   0&   0&   137\\
$hhhc$& 23&  21&  25&  20&   0&   0&   0&   0&   0&   0&   0&   2&   2&   3&   1&   0&   0&  4464\\
$hhhe$& 29&  30&  15&  11&   0&   0&   0&   0&   0&   0&   0&   4&   4&   2&   2&   0&   0&   137\\
$hhhh$& 21&  11&  60&   4&   0&   0&   0&   0&   0&   0&   0&   0&   1&   0&   0&   0&   0& 31327\\
\hline\end{tabular}}\smallskip

\subsection{The discrete structural states}

The marginal one-dimensional distribution of the pseudobond
bending angle has two prominent peaks around $\theta =1.10$ and 1.55
(radians). Non-zero $\theta$s are in the interval $[.4, 1.9]$.
The marginal one-dimensional distribution of the torsion angle
$\tau$ has one immediately noticeable peak at $\tau =0.87$
(corresponding to the helix).
Another peak at $\tau =-2.94$ is less prominent. There is a vague
peak still recognizable around $\tau =-2.00$. A grid generated
with $\theta \in \{1.00, 1.55\}$ and $\tau\in \{ -2.80, -2.05,
-1.00, 0.00, 0.87\}$ is used to search high dimensional phase
space for density peaks by the downhill simplex method. In the box
counting, the box size is taken from $0.1$ to $0.2$ for $\theta$,
and the width for $\tau$ is twice of that for $\theta$. The helices
seen as a single peak in the three-angle phase space are clearly
identified as several sub-peaks in the five-angle phase space.
Further exploring main peaks in the five-angle phase space, we
identify 17 mode centers, which are then used as the main initial
parameters to train the mixture model. Finally, the 17 structural
states are obtained for the mixture model by the EM algorithm.
They are listed in Table 1. Note that it is the entries of the
inverse covariance matrix that are given. The determinant of the
matrix is a measure of the divergence of the corresponding mode.
The most sharp state is $H$, while the most vague state is
$Q$, which occupies the least proportion of phase points.

\begin{center}%}\end{center}\smallskip
\parbox{15cm}{Table 3. The forward transition rates (multiplied by 100)
between structural states.\\
}\end{center}\vspace{-.5cm}

\begin{tabular}{c rrrrrrrrrrrrrrrrr}\hline
  & $I$&$J$&$H$&$K$&$F$&$E$&$C$&$D$&$A$&$B$&$G$&$L$&$M$&$N$&$O$&$P$&$Q$\\
\hline
$I$& 28& 13& 26& 12&  0&  0&  0&  0&  0&  0&  8&  3&  6&  0&  4&  0&  0\\
$J$& 19& 21& 23& 12&  0&  0&  0&  0&  0&  0& 11&  3&  5&  0&  5&  0&  0\\
$H$& 11&  7& 74&  2&  0&  0&  0&  0&  0&  0&  2&  0&  1&  0&  3&  0&  0\\
$K$&  8&  6&  4& 23&  2&  3&  2&  0&  5&  2&  8& 14& 11&  4&  4&  6&  0\\
$F$&  0&  0&  0&  0&  3& 32& 16& 16& 14&  8&  0&  1&  0&  5&  0&  1&  5\\
$E$&  0&  0&  0&  0&  6& 44& 12& 15&  5&  7&  0&  1&  0&  3&  0&  4&  2\\
$C$&  0&  0&  0&  3&  7& 22& 22&  0& 17&  5&  1&  3&  1& 10&  0&  8&  1\\
$D$&  0&  0&  0&  0&  7& 44& 14&  8& 10&  6&  0&  2&  0&  3&  0&  4&  2\\
$A$& 12& 26& 19&  8&  1&  0&  0&  0&  0&  0& 13& 10&  5&  0&  2&  2&  0\\
$B$&  2&  2&  1&  5&  6& 12&  5&  0&  2&  4&  9& 28&  8&  4&  2& 10&  1\\
$G$&  0&  0&  0&  1&  7& 21& 28&  7& 11&  7&  0&  2&  1&  6&  0&  7&  2\\
$L$&  0&  0&  0&  0&  2& 20& 17& 22& 12&  9&  0&  1&  0&  5&  0&  2&  9\\
$M$& 14& 10&  8& 11&  2&  3&  2&  0&  2&  2& 10& 13&  9&  3&  3&  8&  0\\
$N$&  3&  5&  2&  4&  3&  2&  2&  0&  1&  1& 35&  8& 13&  3& 13&  4&  0\\
$O$&  1&  1&  0&  2&  2&  2&  1&  0&  0&  3& 56&  6& 17&  2&  4&  4&  0\\
$P$& 12& 18&  7&  7&  2&  1&  1&  0&  0&  0& 12& 23&  9&  1&  4&  3&  0\\
$Q$&  0&  0&  0&  1&  3& 16& 15&  8& 10&  5&  1&  4&  1& 20&  0&  3& 13\\
\hline\end{tabular}\smallskip \\

\subsection{The structure alphabet and the secondary structure}
Our 17 structural states or letters of the structural alphabet
describe the local structure of
four-residue segments, and a code is assigned to the third residue
of the unit of four residues. The total number of possible
four-residue secondary structures is 37. The restriction of the
minimal lengths 2 for $e$ and 3 for $h$ removes 44 quartets from
the total $3^4=81$.

In order to make a detailed comparison between the secondary
structures and the discrete structural states, from the training
set we extract a subset, which contains 676 fragments and 118,621
residues (hence 116,593 points in the three-angle phase space). We
arrange the corresponding counts in Table~2. (Secondary
structure $heeh$ has zero count, so it is omitted.) The table shows
the percentages of each secondary structure in the structural
states. It is clearly seen that there
exists a correlation between the two types of structure
classifications. For example, from Table.~2 $hhhh$ are mainly
attributed to $H$, $I$ and $J$, while $eeee$ to $E$, and $D$.
The mutual information between the conformational codes and the
secondary structure states equals $0.731$. In Table 2, %0.730728
the row $cccc$ shows rather uniform percentages in different
structural states as we would expect.

\subsection{Transition between structural states}
Any two sequential points $(\theta_{i-1} ,\tau_i ,\theta_i)$ and
$(\theta_i, \tau_{i+1}, \theta_{i+1})$ share the common angle
$\theta_i$. The effect of the connection of sequential structural
states reflects transition rates between structural states. We
first convert the 3D structures of the training set to their
structure code sequences, and then determine the transition rates
by counting code pairs. The obtained rates are listed in Table 3.

The entries of Table 3 are the forward transition rates as the
conditional probability of the $(i+1)$-th site of a state chain at
a given $i$-th site. Normalized according to the row, the table
tells where a row state would like to go. Extended states, e.g.
$H$ and $E$, are characterized by large diagonal elements, while
transient states, e.g. $A$ and $G$, have almost vanishing diagonal
rates. From the table, we may trace the capping states
for the helix and $\beta$-strand. For example, $A$ is an important mode
which leads to the helix, and $G$ is a main leaving mode for the helix.

\begin{center}%}\end{center}\smallskip
\parbox{15cm}{Table 4. CLESUM: The conformation letter substitution
matrix (in the unit of 0.05 bit).\\
}\end{center}\vspace{-.5cm}

{\small\begin{tabular}{c rrrrrrrrrrrrrrrrr}\hline
J& 38&    &    &   &   &   &   &   &   &   &   &   &   &   &  &  &   \\
H& 15&  25&    &   &   &   &   &   &   &   &   &   &   &   &  &  &   \\
I& 12&  14&  51&   &   &   &   &   &   &   &   &   &   &   &  &  &   \\
K& 16&   8&  17& 51&   &   &   &   &   &   &   &   &   &   &  &  &   \\
N& -1& -32& -16& 28& 89&   &   &   &   &   &   &   &   &   &  &  &   \\
Q&-43& -87& -69&-24& 31& 88&   &   &   &   &   &   &   &   &  &  &   \\
L&-31& -61& -48&  0&  5& 24& 71&   &   &   &   &   &   &   &  &  &   \\
G&-21& -49& -40&-11& -7&  8& 27& 68&   &   &   &   &   &   &  &  &   \\
M& 17&  -2&  -4& 14&  8& -7&  4& 21& 59&   &   &   &   &   &  &  &   \\
B&-55& -94& -79&-49&-11& 10&-13& 12&-14& 49&   &   &   &   &  &  &   \\
P&-33& -58& -55&-35& -4&  6&-14&  3&  7& 41& 64&   &   &   &  &  &   \\
A&-22& -43& -39&-17& 10& 13&-12& -7& -2& 19& 34& 71&   &   &  &  &   \\
O&-23& -54& -37&  5& 14&-13& -5& -2&  5&-12&  2& 23&102&   &  &  &   \\
C&-42& -75& -59&-32& -5& 27& -2& -6&-12&  5&  4& 12&  1&51 &  &  &   \\
E&-91&-125&-112&-83&-43& -8&-23&-24&-47& 13& -6&-27&-49& 2 &34&  &   \\
F&-73&-106& -95&-67&-32&  0&-18& -6&-34&  4& -2&-22&-31&19 &24&48&   \\
D&-87&-122&-105&-81&-45& 13&-24&-32&-50& 11&-11&-19&-43&19 &21&20& 49\\
 & J &  H &   I& K & N & Q & L & G & M & B & P & A & O &C  &E &F & D \\
\hline\end{tabular}}\smallskip\\

\subsection{Structural substitution matrix}

Sequence alignment is the main procedure of comparing sequences.
Certain amino acid substitutions commonly occur in related
proteins from different or same species. Amino acid substitution matrices,
extracted from our knowledge of most and least common changes
in a large number of proteins, serve for the purpose of
sequence alignment. The popular BLOSUM matrix of Henikoff and
Henikoff (1992) is derived from a large set of conserved amino
acid patterns without gaps representing various families. The
frequency of amino acid substitutions in alignments is counted in
sequence alignments. These frequencies are then divided by the
expected frequency of finding the amino acids together in an
alignment by chance. The ratio of the observed to the expected counts
is an odds score. The BLOSUM entries are logarithms of the odds
scores with the base 2 and multiplied by a scaling factor of 2.

To use our structural codes directly for the structural
comparison, a score matrix similar to BLOSUM is desired. There is
a database of aligned structures, the FSSP of Holm and Sander
(1997), which is based on exhaustive all-against-all 3D structure
comparison of protein structures in the PDB.
The proteins in the FSSP are divided into a representative set and
sequence homologs of the representative set. The
representative set contains no pair which have more than 25\%
sequence identity. In the version of Oct 2001, there are 2,860
sequence families representing 27,181 protein structures. A tree
for the fold classification of the representative set is
constructed by a hierarchical clustering method based on the
structural similarities. Family indices of the FSSP are obtained
by cutting the tree at levels of 2, 4, 8, 16, 32 and 64 standard
deviations above database average. We convert the structures of the
representative set to their structural code sequences. All
the pair alignments of the FSSP for the proteins with the same
first three family indices in the representative set are collected
for counting aligned pairs of structural codes. The total
number of code pairs are 1,143,911. The substitution
matrix derived in the same way as the BLOSUM was obtained is
shown in Table 4, where a scaling factor of 20 instead
of 2 is used to show more details. We call this conformation
letter substitution matrix
CLESUM. Henikoff and Henikoff (1992) introduced for their BLOSUM
the average mutual information per amino acid pair $H$, which is
the Kullback-Leibler distance between the joint model of the
alignment and the independent model. The value of $H$ for our
CLESUM equals $1.05$, which is close to that for BLOSUM83.

\section{Discussion}
Biologically important modules have been repeatedly employed in
protein evolution by gene duplication and rearrangement
mechanisms. They form components of fundamental units of structure
and function. The presence of modules provides a guide to classify
proteins into module-based families, and helps the structure
prediction. The existence of such conservative recurrent segments
sets a solid foundation for the local analysis. We have
discretized the combination of three psuedobond angles formed by
four consecutive $C_\alpha$ atoms to convert the local geometry to
17 coarse-grained conformational letters according to a mixture
model of the angle distribution.

\subsection{The precision of the conformational codes}
From the correlation between the conformational codes and the secondary
structures (Table 2), it is not surprising that there exists a propensity
of the codes to amino acids. The coarse-graining would introduce an error.
It is then important to examine the precision of the codes. For this purpose,
we randomly pick up 1,000 points for each code, and calculate the distance
root mean squared deviation ($drms$) for each of the total 499,500 pairs
from their coordinates. The $drms$ of
structures $a$ and $b$, without requiring a structure alignment, is defined
as the averaged distance pair difference
\begin{equation}
drms = \left[\frac 2{n(n-1)}\sum_{i=2}^n\sum_{j=1}^{i-1}(|{\bf
r}_{ai}- {\bf r}_{aj}|-|{\bf r}_{bi}-{\bf
r}_{bj}|)^2\right]^{1/2},
\end{equation}
where ${\bf r}_{ai}$ is the coordinate of atom $i$ in structure
$a$. The averaged coordinate pair difference, i.e. the coordinate
root mean squared deviation $crms$, is about $1.2$ times of the
$drms$.

The errors of the conformational codes are listed in Table~5. The most
precise code $H$ has an error $0.133\pm 0.060$\AA ,while the vaguest code
$L$ has an error $0.604\pm 0.365$\AA . After averaging over the code
relative frequencies, the mean error is $0.330$\AA .

\begin{center}
\parbox{15cm}{Table 5. The errors of the conformational codes.\\
}\end{center}\vspace{-.5cm}{\small

\begin{tabular}{c rrrrrrrrr}\hline
Conformational code      &I\phantom{a.}&J\phantom{a.}&H\phantom{a.}
&K\phantom{a.}&F\phantom{a.}&E\phantom{a.}&C\phantom{a.}&D\phantom{a.}\\
Mean drms (\AA )&0.244&0.246&0.133&0.452&0.398&0.307&0.392&0.262\\
Standard deviation (\AA )&0.110&0.124&0.060&0.219&0.287&0.173&0.218&0.149\\
\hline
\multicolumn{1}{r}{A\phantom{a.}}&B\phantom{a.}&G\phantom{a.}&L\phantom{a.}
&M\phantom{a.}& N\phantom{a.}& O\phantom{a.}& P\phantom{a.}& Q\phantom{a.}\\
\multicolumn{1}{r}{0.347}&0.322&0.390& 0.604& 0.481& 0.551& 0.538& 0.252&0.506\\
\multicolumn{1}{r}{0.163}&0.197&0.192&0.365&0.231&0.321&0.318&0.134&0.287\\
\hline\end{tabular}}

\subsection{The connection effect of sequential states}
Compared with the HMM, the mixture model does not include the
connection effect of sequential states. The parameter number
increases quadratically with the number of categories for a Markov
model, while only linearly for a mixture model. We have to
compromise between precision and correlation. A mixture model with
fine categories is also promising.

Since the model training involves a global optimization the choice
of a good initial trial plays an important role. A careful
exploration of the density distribution in the five-angle space
corresponding to two consecutive conformational states reveals
that the peaks in the five-angle space give a finer picture of the
peaks in the three-angle space. That is, subpeaks in the
three-angle space are easily recognizable from peaks in the
five-angle space. We have identified 17 intense peaks, which
survive the later process of model training. Camproux et al.
(1999) found 12 modes for the four-residue unit by a HMM. Instead
of angles, they used a combination of four distances. Since only
three of the four are independent their mode centers need not
correspond to a real conformation. However, we still can see the
correspondence between their codes and ours: $\alpha_1$-$H$,
$\alpha_2$-$J$, $\alpha '$-$K$, $\alpha '_-$-$O$, $\alpha
'_+$-$M$, $\gamma_1$-$N$, $\gamma_2$-$P$, $\gamma_\beta$-$Q$,
$\gamma_{\beta\alpha}$-$A$, $\gamma_{\alpha\beta}$-$G$,
$\beta_2$-$D$, and $\beta_1$-$E$.

\subsection{Structure alignment via conformational codes}
The conversion of a 3D structure of coordinates to its
conformational codes requires little computation. To distinguish
from the amino acid sequence, we call the converted code sequence
the code series, or simply series. Once we transform 3D structures
to 1D series, the structure comparison becomes the series
comparison. Tools for analyzing ordinary sequences can be directly
applied. We have constructed the conformational letter
substitution matrix CLESUM from the alignments of the FSSP
database. We shall examine the performance of the conformational
alphabet derived above.

\begin{center}
\parbox{15cm}{Table 6. The alignment of 1urnA and 1ha1. The first two
lines are their amino acid sequences aligned according to the
FSSP, while the last two lines are the global Needleman-Wunsch
alignment of the conformational code series. Lowercase letters of
amino acids indicate
structural nonequivalence.\\
}\end{center}\vspace{-.8cm} \hspace{3cm}\vbox{
\begin{verbatim}
1urnA avpetRPNHTIYINNLNEKIKKDELKKSLHAIFSRFGQILDILVSRS
1ha1b    ahLTVKKIFVGGIKEDT    EEHHLRDYFEQYGKIEVIEIMTDRGS
        CCPMCEALEEEENGCPJGCCIHHHHHHHHIKMJILQEPLDEEEBGAIK
        ...BBEBGEDEENMFNMLFA....HHHHHKKMJJLCEBLDEBCECAKK

1urnA LKMRGQAFVIFKEVSSATNALRSMqGFPFYDKPMRIQYAKTDSDIIAKM
1ha1b GKKRGFAFVTFDDHDSVDKIVIQ kYHTVNGHNCEVRKAL
      ...GNGEDBEEALAJHHHHHHIKKGNGCENOGCCEFECCALCCAHIJH
      AGCPOLEDEEEALBJHHHHI.IJGALEEENOGBFDEECC.........
\end{verbatim}}

Holm and Sander (1998) gave an example of the
$\alpha$/$\beta$-meander cluster with four members showing
different levels of structural similarity. Their PDB-IDs are
1urnA, 1ha1, 2bopA and 1mli. The structure of 1urnA was taken as
the frame to superimpose the other structures. The structural
similarity to 1urnA from high to low are 1ha1, 2bopA and 1mli.
Taking the scaling factor for the CLESUM to be 2, and using $-12$
for the the gap-opening penalty and $-4$ for the gap extension,
the global Needleman-Wunsch alignment of 1urnA and 1ha1 is shown
in Table 6, where, in the first two lines, the amino acid
sequences aligned according to the FSSP are also given. It is seen
that, except for segment boundaries, the two alignments coincide.
The alignment of the FSSP and the code series alignment for 1urnA and
2bopA have three common segments falling in positive score regions
of the series alignment. In the alignments for 1urnA and 1mli two
common segments longer than 8 are still seen. As for the amino acid
sequence alignment, in the case of 1urnA and 1ha1 two segments of
lengths 13 and 21 of the sequence alignment coincide with the FSSP,
but no coincidence are seen in the other two cases.

The conformational codes are local. Even though a global alignment
algorithm is used, this does not guarantee that the found alignment
corresponds to the optimal structure superposition. However, the code
series alignment does not affected by the domain move, is then good
for analyzing the structure evolution. For example, the first helix of 1ha1 is
shorter than its counterpart in 1urnA by one turn. The FSSP aligns
the $N$-cap (with codes $FA$) of the 1ha1 helix to the helix
(with codes $HH$) of 1urnA, but local structure $FA$ is closer
to $CC$ (with positive scores) than to $HH$
(with negative scores).% and $drms =0.84$\AA).

The CLESUM includes only the structural information. When we
compare two structures we usually know also their amino acid
sequences. Many papers considered a linear combination of
structural alignment score and sequence alignment score. This is
an approximation of independency. From the FSSP, it is possible to
construct a substitution matrix in the joint space of the
structure and sequence. However, such a matrix would have about
$6\times 10^4$ parameters. When the structure is to be emphasized,
we may use a reduced amino acid alphabet (Zheng, 2004). For example, clustering
20 amino acids into 3 groups would reduce the parameter number to
about $10^3$. We often want to compare a sequence with unknown
structure to a known structure. In this case, a rectangular
substitution matrix of the type of (amino
acid)$\times$(conformational code) to (amino acid) is useful. The
construction of these matrices is our next task.

It is known that the sequence-structure relationships have not
always been strong. Bystroff and Baker (1998) have built a library
of structure-sequence motifs, which are expected to correspond to
functional units recurring in different protein contexts and to be
found in different combinations in distantly related or
functionally unrelated proteins. To identify the structural
features that have strong sequence preferences is to locate peaks
of density distribution in the joint structure-sequence space.
Previously, the structure-based clustering was a duty much heavier
than the sequence-based clustering, so one had to start with a
sequence-based clustering, and was kept constantly to run between
the structure and sequence subspaces. It is then interesting to
see whether the library can be improved by clustering directly in
the joint structure-sequence space with the help of conformational
codes. This is under study.

\begin{quotation}
{This work was supported in part by the Special Funds for Major
National Basic Research Project and the National
Natural Science Foundation of China.}
\end{quotation}

%\newpage

\end{document}